\documentclass[a4paper]{article}
\usepackage{INTERSPEECH2020}
\usepackage[dvipsnames]{xcolor}
\usepackage{amsmath}
\usepackage{multirow}
\usepackage{microtype}
\usepackage{cite}
\usepackage{booktabs}

\usepackage{xcolor}
\usepackage{multirow}
\usepackage{comment}
\usepackage{url}

\usepackage[normalem]{ulem}

\title{Robust Self-Supervised Audio-Visual Speech Recognition}
\name{Bowen Shi$^{1*}$\thanks{$^*$Work done at Meta AI}, Wei-Ning Hsu$^2$, Abdelrahman Mohamed$^2$}

\address{
  $^1$Toyota Technological Institute at Chicago\\
  $^2$Meta AI}
\email{bshi@ttic.edu, wnhsu@fb.com, abdo@fb.com}

\begin{document}

\maketitle
\begin{abstract}
Audio-based automatic speech recognition (ASR) degrades significantly in noisy environments and is particularly vulnerable to interfering speech, as the model cannot determine which speaker to transcribe. Audio-visual speech recognition (AVSR) systems improve robustness by complementing the audio stream with the visual information that is invariant to noise and helps the model focus on the desired speaker. However, previous AVSR work focused solely on the supervised learning setup; hence the progress was hindered by the amount of labeled data available. 
In this work, we present a self-supervised AVSR framework built upon Audio-Visual HuBERT (AV-HuBERT), a state-of-the-art audio-visual speech representation learning model. On the largest available AVSR benchmark dataset LRS3, our approach outperforms prior state-of-the-art by $\sim 50\%$ (28.0\% vs. 14.1\%) using less than 10\% of labeled data (433hr vs. 30hr) in the presence of babble noise, while reducing the WER of an audio-based model by over 75\% (25.8\% vs. 5.8\%) on average \footnote{Our code and models are available at \url{https://github.com/facebookresearch/av_hubert}}.
\end{abstract}

\section{Introduction}
\label{sec:intro}
With the recent development of supervised neural models~\cite{hinton_2012,amodei16deepspeech}, the performance of automatic speech recognition (ASR) systems has improved significantly, achieving human parity~\cite{xiong2016humanparity} or even outperforming humans on several clean speech benchmarks~\cite{Tske2020SingleHA,Nguyen2021SuperHumanPI}. However, ASR systems are vulnerable to noise and may degrade drastically when speech recordings are corrupted with noise~\cite{Afouras2018TheCD}. To make ASR more reliable in various scenarios, research on noise robustness~\cite{Watanabe2020CHiME6CT,Vincent2017AnAO,Kinoshita2020ImprovingNR} has received increasing attention in recent years.

An active research direction on noise robustness combines the audio and visual streams of the speaker to utilize the noise-invariant lip movement information. Audio-visual speech recognition (AVSR) models, which combine these two modalities, bring AI systems one step closer to how humans perceive speech~\cite{mcgurk1976hearinglips} and provide better performance for a broad range of application scenarios~\cite{Biswas2016camera,Koguchi2018AMC} where both audio and visual streams are accessible, e.g., video meetings, talks, interviews.

Although early studies of audio-visual speech recognition (AVSR) appeared more than 60 years ago~\cite{Sumby1954VisualCT}, recent developments on novel model architectures~\cite{afouras2018deepavsr,Xu2020DiscriminativeMS} and large-scale data collection~\cite{afouras2018deepavsr,afouras2018lrs3} have brought AVSR performance to new heights. Nonetheless, while modern neural architectures are hungry for large training data, existing research AVSR efforts are fully-supervised, requiring costly labeled data. This limitation hinders the application of modern AVSR systems in low-resource settings, which is the case for most of the $\sim$7,000 spoken languages~\cite{ethnologue}.

This paper presents a self-supervised framework for robust AVSR, which is based upon the recently introduced Audio-Visual HuBERT (AV-HuBERT) pre-training appraoch~\cite{avhubert}. First, large quantities of unlabeled audio-visual speech data are used to pre-train our model to capture the nuanced correlations between sounds and associated lip movements, then only a tiny amount of transcribed audio-visual speech data is used for fine-tuning the model for best AVSR performance. The efficacy of our framework is demonstrated on low-resource (30h) and mid-resource (433h) setups showing WER reductions of up to 50\% compared to previous SOTA models.
Furthermore, we investigate the robustness of the proposed approach and audio-only systems against different types of noises, which have not been studied in prior work but are essential for practical applications. For example, an AVSR system deployed in meeting scenarios is subject to babble noise, while one used in a home environment naturally encounters music, cooking, or vacuums machine noises.
\section{Method}
\label{sec:method}

In this section, we present our methodology for audio-visual speech recognition. First, we introduce the Audio-Visual HuBERT (AV-HuBERT) pre-training approach, which we use for unsupervised learning of joint representations over audio and visual streams. We then describe how we adopt AV-HuBERT for robust audio-visual speech recognition.

\subsection{AV-HuBERT for Audio-Visual Speech Recognition}
\label{sec:av-hubert}

\begin{figure}[htp]
    \centering
    \caption{AV-HuBERT for audio-visual speech recognition. \textcolor{red}{X}: mask; blue waveform: original audio; orange waveform: noise; $C_n$: audio-visual clusters. Dashed box: the pre-trained part}
    \includegraphics[width=\linewidth]{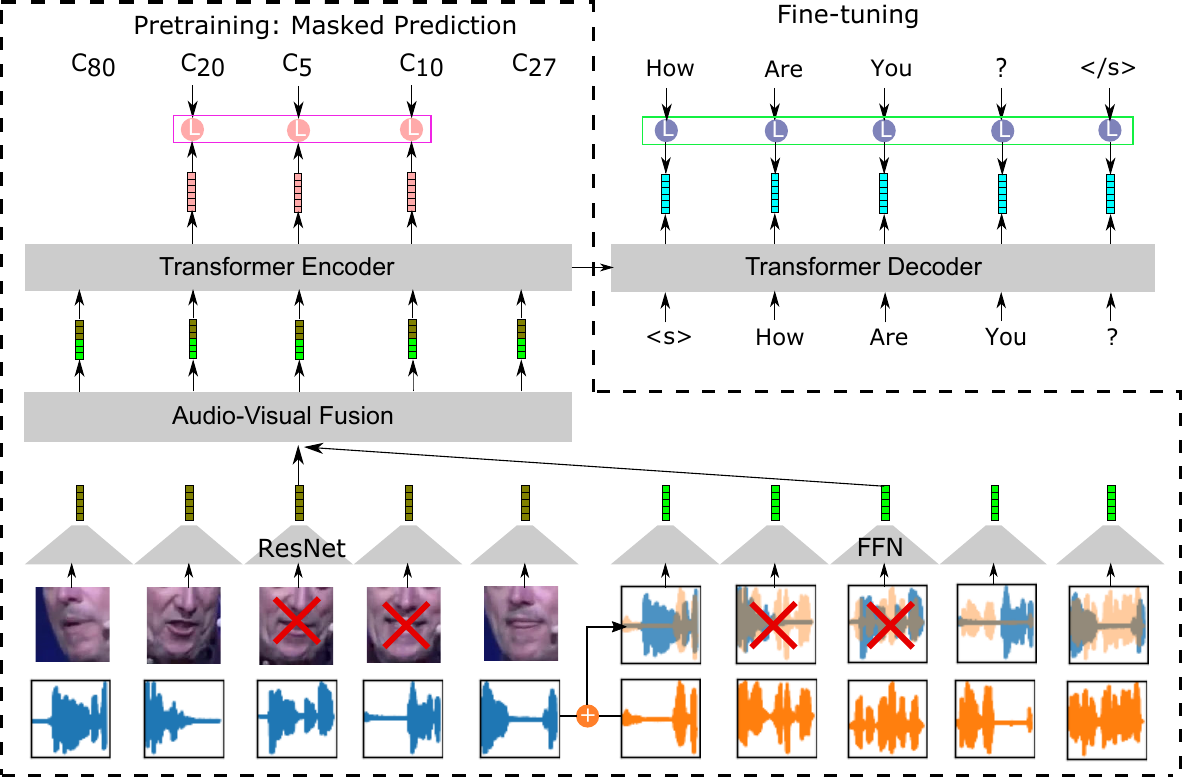}
\end{figure}

AV-HuBERT~\cite{avhubert} is a self-supervised approach for learning joint speech representations from audio and lip-movement information in video recordings, which extends the HuBERT~\cite{Hsu2021HuBERT} speech representation learning framework to multimodal inputs. AV-HuBERT consumes frame-level synchronous audio and video streams as input to produce contextualized audio-visual representations for each frame. AV-HuBERT pretraining iterates over two steps: feature clustering and masked prediction. 

Feature clustering creates discrete frame-level targets for the subsequent masked prediction step. Audio-based mel-frequency cepstral coefficients (MFCC) features are always used for cluster generation in the first iteration. For multi-iteration pretraining, the learned audio-visual features extracted from the latest AV-HuBERT transformer network are used for cluster generation in all subsequent iterations. Inspired by the BERT pretraining widely used for text data~\cite{bert} and deep cluster for visual data~\cite{caron2018deep}, The masked prediction loss drives training of the AV-HuBERT model by predicting the cluster assignments of the masked frames given a corrupted video signal with randomly masked segments. To finetune it for a downstream task, the cluster prediction head of the pretrained model is removed. Depending on the desired architecture of the final model, either a linear layer is added for an encoder-only model or a randomly initialized decoder module with cross attention over the pretrained encoder is used for a sequence-to-sequence model. Some or all layers may be updated during finetuning.

Unlike the prior work in ~\cite{avhubert}, which utilizes the pretrained AV-HuBERT encoder for unimodal downstream scenarios like lip-reading and ASR, this paper examines the effectiveness of the multimodal learned representations of AV-HuBERT for the multimodal audio-visual speech recognition (AVSR) task that aims to transcribe speech videos using audio and visual streams. Given a pre-trained AV-HuBERT model, we keep both its audio and video frontends during finetuning. We use a sequence-to-sequence model for AVSR, where AV-HuBERT serves as the encoder module. In contrast to pretraining, we do not apply input masking or modality dropout during finetuning. Also, we froze the pretrained AV-HuBERT encoder for a certain number of training steps, after which we updated all model weights.

\subsection{Noise-Augmented AV-HuBERT}
AVSR systems leverage the visual modality during noisy conditions~\cite{afouras2018deepavsr}; however, A recognizer trained on clean conditions may rely overly on the audio stream since a model can predict with audio more effortlessly, thus leading to failure of leveraging visual information in adverse auditory conditions at test time. A typical solution adopted by prior work is noise-augmented supervised training~\cite{afouras2018deepavsr,Xu2020DiscriminativeMS}, which adds noise sampled from a separate noise dataset to the clean audio at a fixed or sampled signal-to-noise ratio (SNR). We adopt this strategy during the finetuning stage and refer to it as \textit{noise-augmented finetuning} to emphasize the stage noise is employed.

To further boost our model's robustness to acoustic noise, we extended noise augmentation to AV-HuBERT pretraining by randomly adding different types of noise to the audio input, making AV-HuBERT more suitable for AVSR applications. We refer to it as \textit{noise-augmented pretraining}. Incorporating noise in the pretraining phase benefits the model by closing the domain gap between pretraining, finetuning, and testing. We still use the cluster assignment inferred from clean audio-visual speech because phonetic information, which is highly correlated with the clusters, should be invariant to noise.

A concurrent work, WavLM~\cite{chen2021wavlm}, proposes utterance mixing, which is a similar technique to ours but applied for audio-only speech representation learning. Utterance mixing augments input audio by randomly sampling speech utterances from the same minibatch. We use more diverse sources in our noise-augmented pretraining, including both speech and non-speech noise, e.g., ambient and babble noise. Additionally, since WavLM targets audio-only self-supervised learning, the intersection between the secondary and the primary utterances needs to be fewer than 50\% to signify which utterance in the mixture is the main one. Our approach is unconstrained and more flexible on mixing noise because the accompanying visual stream disambiguates the primary and secondary utterances.
\section{Experiments}
\label{sec:exp}

\subsection{Data and Experimental Setup}
\label{sec:exp-data-setup}
Our experiments are conducted on LRS3~\cite{afouras2018lrs3} with around 433 hours of audio-visual speech from over 5000 speakers, which is the largest publicly available labeled audio-visual speech recognition dataset. VoxCeleb2~
\cite{voxceleb2}, a large-scale audio-visual speech dataset that was initially proposed for the speaker recognition task is used for our self-supervised pre-training. VoxCeleb2 has around 2,442 hours of videos from over 6,000 speakers and contains utterances from multiple languages. We follow the preprocessing steps in~\cite{avhubert} to select the "English" portion, which amounts to 1,326 hours of videos.

We augment input samples using many noise categories. The total duration of noise in each category is shown in table~\ref{tab:noise-amount}. The noise audio clips in the categories of "\texttt{natural}", "\texttt{music}" and "\texttt{babble}" are sampled from MUSAN dataset~\cite{Snyder2015MUSANAM}, while the overlapping "\texttt{speech}" noise samples are drawn from LRS3. In creating "\texttt{speech}" and "\texttt{babble}" noise sets, we ensured there are no speaker overlap among different partitions.

\begin{table}[htp]
    \centering
    \caption{\label{tab:noise-amount}Total duration in hours of noise samples in different categories}
    \begin{tabular}{ccccc}
    \toprule
    Partition & natural & music & babble & speech \\
    \midrule
        train & 6 & 35 & 20 & 50 \\
        validation & 1 & 4 & 2 & 6\\
        test & 1 & 4 & 2 & 6\\
        \bottomrule
    \end{tabular}
\end{table}

We follow the protocol of~\cite{avhubert} to create two settings for finetuning the model; a low-resource setting using 30h of labeled videos and a mid-resource setting using 433h of labels. Unless otherwise specified, we use AV-HuBERT LARGE as the default model architecture for all our experiments. The model has 24 transformer blocks, where each block has 16 attention heads and 1024/4096 embedding/feedforward dimensions. We add a 9-layer randomly initialized transformer decoder with similar embedding/feedforward dimensions during finetuning.

During training, we first select one noise category and sample a noise audio clip from its training partition. We randomly mix the sampled noise at 0dB SNR with a probability of 0.25, following~\cite{Xu2020DiscriminativeMS}. At test time, we evaluate the model separately for each noise type. The testing noise clips are added at five SNR levels: $\{-10, -5, 0, 5, 10\}dB$. The performance on the original clean test set is also reported for comparison. 
By default, noise clips are added during both pre-training and finetuning. We follow the training pipeline in~\cite{avhubert}, where the model is trained for five iterations in total. To save the computation time, we always use the smaller BASE model architecutre~\cite{avhubert} in all iterations except the last one, where we use a LARGE model. Video samples are batched together not to exceed 1000 image frames per GPU. The model is pre-trained with 600K steps using 64 V100-GPUs and finetuned for 30K/100K steps respectively in 30h/433h setting.

\begin{table*}[htb]
\centering
\caption{WER (\%) of our models and prior work on the LRS3 dataset. ``Mode'' denotes whether a model uses audio-visual input (AV) or only audio as input (A). ``Hr'' denotes the amount of labeled audio-visual speech data used in each system.
}
\label{tab:main_avsr}
\resizebox{\linewidth}{!}{
\begin{tabular}{
    lll | 
    cccccc | 
    cccccc | 
    cccccc | 
    c}
\toprule
\multirow{3}{*}{Model} & 
\multirow{3}{*}{Mode} & 
\multirow{3}{*}{Hr} & 
\\
& & &
    \multicolumn{6}{c|}{Babble, SNR=} & 
    \multicolumn{6}{c|}{Speech, SNR=} & 
    \multicolumn{6}{c|}{Music+Natural, SNR=} & 
    Clean \\
& & &
    -10 & -5 & 0 & 5 & 10 & avg &
    -10 & -5 & 0 & 5 & 10 & avg &
    -10 & -5 & 0 & 5 & 10 & avg &
    $\infty$ \\
\midrule
Makino et al.~\cite{Makino2019rnnt} & AV & 31K &
    - & - & - & - & - & - &
    - & - & - & - & - & - & 
    - & - & - & - & - & - &
    4.5 \\
Ma et al.~\cite{ma2021conformer} & AV & 595 &
    - & - & - & - & - & - &
    - & - & - & - & - & - & 
    - & - & - & - & - & - &
    2.3 \\
Afouras et al.~\cite{afouras2018deepavsr} & AV & 1.4K &
    - & - & 42.5 & - & - & - &
    - & - & - & - & - & - & 
    - & - & - & - & - & - &
    7.2 \\
Xu et al.~\cite{Xu2020DiscriminativeMS} & AV & 433 &
    38.6 & 31.1 & 25.5 & 24.3 & 20.7 & 28.0 &
    - & - & - & - & - & - &
    - & - & - & - & - & - &
    6.8 \\
\midrule
AV-HuBERT & AV & 30 &
    35.1 & 18.4 & 8.3 & 4.9 & 3.9 & 14.1 & 
    11.5 & 6.8 & 5.0 & 4.2 & 3.9 & 6.3 & 
    12.0 & 7.0 & 4.8 & 4.1 & 3.7 & 6.3 & 
    3.3 \\
AV-HuBERT & AV & 433 &
    34.9 & 16.6 & 5.8 & 2.6 & 2.0 & 12.4 & 
    11.4 & 4.6 & 2.9 & 2.2 & 1.8 & 4.6 & 
    9.7 & 4.77 & 2.5 & 1.9 & 1.8 & 4.1 & 
    1.4 \\
\midrule
AV-HuBERT & A & 30 &
    99.6 & 69.3 & 21.9 & 9.0 & 5.6 & 41.1 & 
    77.3 & 51.2 & 32.0 & 19. & 10.8 & 38.2 & 
    47.9 & 21.5 & 9.2 & 5.9 & 4.8 & 17.8 & 
    3.8 \\
AV-HuBERT & A & 433 &
    97.5 & 62.3 & 15.7 & 5.1 & 2.6 & 36.6 & 
    81.7 & 56.2 & 37.3 & 19.0 & 8.3 & 40.5 & 
    38.7 & 15.1 & 5.7 & 3.1 & 2.3 & 13.0 &
    1.6 \\
\bottomrule
\end{tabular}
}
\end{table*}

\subsection{Main Results}
Table~\ref{tab:main_avsr} compares the performance of our proposed noise-augmented AV-HuBERT approach under different settings versus existing supervised AVSR models. In the clean audio setting, our best model outperforms the best model from Ma et al.~\cite{ma2021conformer} by ~39.0\% (2.3\%$\rightarrow$1.4\%) while using fewer labeled data. To enable direct comparison with previous research work which focus primarily on \texttt{babble} noise, we follow \cite{Xu2020DiscriminativeMS} to synthesize \texttt{babble} noise by randomly mixing 30 audio clips from LRS3\footnote{\cite{afouras2018deepavsr} uses audios from LRS2~\cite{afouras2018deepavsr}, which has restricted access and we are unable to obtain it.} for evaluation.

As shown in the "\texttt{Babble}" column, with only 30 hours of labeled data, our model outperforms \cite{afouras2018deepavsr} by 80.4\% (42.5\%$\rightarrow$8.3\%) and \cite{Xu2020DiscriminativeMS} by 67.4\% (25.5\%$\rightarrow$8.3\%) at 0dB SNR. Compared to the former SOTA~\cite{Xu2020DiscriminativeMS}, we achieve 49.6\% lower WER (28.0\% $\rightarrow$ 14.1\%) on average across different SNR ratios with 10 times fewer labels. When using all 433 hours labeled data for finetuning, the relative improvement is further increased to 55.7\% (28.0\% $\rightarrow$ 12.4\%). Note that the \texttt{babble} noise we use for training our model is synthesized from MUSAN, which has potential domain mismatch from the \texttt{babble} noise synthesized from LRS3 used at test time; however, our approach significantly improves over prior work and estableshes the new SOTA.

When the noise type is extended beyond \texttt{babble} noise, our proposed audio-visual model consistently improves over its audio-only ASR counterpart with is more than $70\%$ relative WER reduction. The reduction varies depending on noise type and SNR; hence, we analyze how the model performs in different noise conditions and how each component of our approach contributes to such improvement.

\subsection{Analysis}
To examine the impact of pre-training (no pre-training, pre-training with clean audio, or noise-augmented pre-training) and input modality (audio or audio-visual), we experimented with the six setups covering the cross product of these conditions. For setups with audio-only input during finetuning, we follow~\cite{avhubert} by replacing the visual features in the pre-trained AV-HuBERT model with a dummy zero vector at each frame. Performances of these six setups are shown in table~\ref{tab:avg-comparison}. Figure~\ref{fig:overall-comparison} shows a more detailed performance breakdown over SNR ratios and noise types.

\subsubsection{Effect of the visual modality}

\begin{figure}[htb]
\caption{\label{fig:overall-comparison}Comparison of models using different inputs and pre-training methods.}
\begin{tabular}{c}
  \includegraphics[width=.95\linewidth]{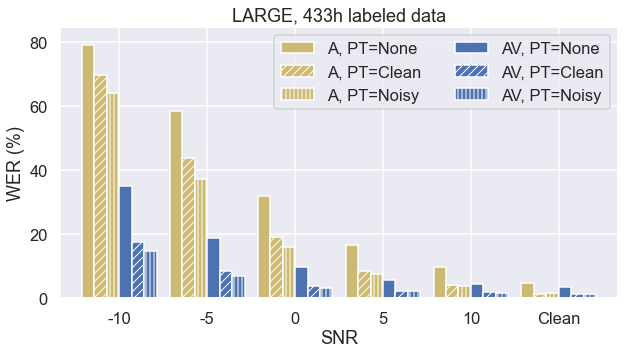}\\ \includegraphics[width=.95\linewidth]{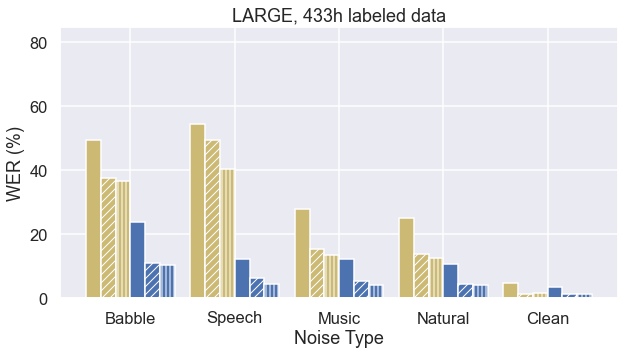}
\end{tabular}
\end{figure}

\begin{table}[htb]
    \centering
    \caption{Comparison among models with different pre-training configurations and input modalities. C: clean audio, N: noisy audio. The N-WER is averaged over 4 noise types and 5 SNRs.}\label{tab:avg-comparison}
    \resizebox{\linewidth}{!}{
    \begin{tabular}{ccc|cc|cc}
        \toprule
        Model   & PT    & FT    & \multicolumn{2}{c|}{Audio-only} & \multicolumn{2}{c}{Audio-visual} \\
        Size    & Type  & Data & C-WER & N-WER & C-WER & N-WER \\
        \midrule
        (a). LARGE & None  & 30h & 20.6 & 59.2 & 20.8 & 42.9 \\
        (b). LARGE & Clean & 30h & 4.3  & 39.8 & 3.3  & 9.3 \\
        (c). LARGE & Noisy & 30h & 3.8  & 28.7 & 3.3  & 7.8 \\
        \midrule
        (d). LARGE & None  & 433h & 4.7 & 39.2 & 3.5 & 14.8 \\
        (e). LARGE & Clean & 433h & 1.5 & 29.1 & 1.4 & 6.9  \\
        (f). LARGE & Noisy & 433h & 1.6 & 25.8 & 1.4 & 5.8  \\
        \bottomrule
    \end{tabular}
    }
\end{table}
We first examine the performance of AVSR models against audio-only ASR models under low-resource and mid-resource conditions by comparing the blue and yellow bars of the same shading pattern in each group in figure~\ref{fig:overall-comparison} and audio-only vs. audio-visual columns in Table~\ref{tab:avg-comparison}. AVSR consistently outperforms audio-only ASR under all settings regardless of the SNR and the type of noise, except for the setup where the model is trained on only 30 hours of labeled data from scratch.

As shown in Figure~\ref{fig:overall-comparison}, the benefit of incorporating the visual stream is more apparent in challenging scenarios, where the WER degradation relative to the clean condition is large. Specifically, these scenarios include low SNR conditions where the volume of the noise is higher and noisy environments with \texttt{speech} or \texttt{babble} noise where the interfering signal is similar to the target speech. Averaged over different pre-training configurations, the AVSR model achieves 53.0\% (42.6\% $\rightarrow$ 20.0\%) and 70.8\% (31.3\% $\rightarrow$ 9.2\%) relative WER reduction over audio-only ASR under noisy settings using 30 hours and 433 hours of labeled data, respectively. 

It is worth noting that our AVSR model enjoys its largest gain over the audio-only model under \texttt{speech} noise settings, where a secondary speech utterance is randomly mixed into the primary one. When using overlapping speech noise under the mid-resource setting, the WER went from 48.1\% to 7.7\% by AVSR on average across different pre-training configurations, while the WER is reduced from 25.8\% to 9.6\% in the other three noise categories. These results suggest that the paired visual stream provides an effective audio source separation, where the audio-only recognizer can not distinguish two audio tracks.

\subsubsection{Effect of pre-training}
\label{sec:pre-train-vs-scratch}
The performance of fine-tuning an AV-HuBERT model is compared against directly optimizing a model from scratch on labeled audio-visual speech in (AV, PT=Clean) vs. (AV, PT=None) bars of Figure~\ref{fig:overall-comparison} and (Clean vs. None) rows (i.e., (b) vs. (a) and (e) vs. (d)) in Table~\ref{tab:avg-comparison}. Note that the AV-HuBERT model pre-trained on clean audio (PT=clean) is identical to the one used in \cite{avhubert}.

On average, AV-HuBERT pre-training brings substantial relative improvements of 78.3\% (42.9\%$\rightarrow$9.3\%) and 53.4\% (14.8\%$\rightarrow$6.9\%) when using 30h and 433h of labeled data, respectively. The model achieves bigger gains in the low-resource setting, which confirms the impact of the self-supervised audio-visual representations learned by the AV-HuBERT model.

\subsubsection{Effect of noise-augmented pre-training}
The impact of incorporating noise in pre-training is presented in (AV, PT=Noisy) vs. (AV, PT=Clean) bars from Figure~\ref{fig:overall-comparison} and (Noisy vs. Clean) rows in Table~\ref{tab:avg-comparison} (i.e., (b) vs. (c) and (e) vs. (f)).

Overall the noise-augmented pre-training improves the result in noisy settings. The WER is reduced by 16.1\% (9.3\%$\rightarrow$7.8\%) / 15.9\% (6.9\%$\rightarrow$5.8\%) on average in low-resource (30h) and mid-resource (433h) settings compared to pre-training on clean data. Compared to an audio-visual model trained from scratch, the noise-augmented pre-training approach reduces recognition error by 81.8\% (42.9\%$\rightarrow$7.8\%) / 60.8\% (14.8\%$\rightarrow$5.8\%) in low/mid-resource settings.

\begin{table}[htb]
    \centering
    \caption{Comparison among BASE models with different pre-training configurations and input modalities. The model is fine-tuned with 30h labeled data. C: clean audio, N: noisy audio. The N-WER is averaged over 4 noise types and 5 SNR ratios.}\label{tab:avg-base-comparison}
    \resizebox{\linewidth}{!}{
    \begin{tabular}{ccc|cc|cc}
        \toprule
        Model   & PT    & FT    & \multicolumn{2}{c|}{Audio-only} & \multicolumn{2}{c}{Audio-visual} \\
        Size    & Type  & Type  & C-WER & N-WER & C-WER & N-WER \\
        \midrule
        (a). BASE & None  & Clean & 24.6 & 79.8 & 22.0 & 70.9 \\
        (b). BASE & Clean & Clean & 4.6  & 46.3 & 4.0  & 28.2 \\
        (c). BASE & Noisy & Clean & 4.4  & 33.8 & 4.1  & 12.5 \\
        \midrule
        (d). BASE & None  & Noisy & 16.9 & 55.4 & 17.2 & 39.5 \\
        (e). BASE & Clean & Noisy & 4.8  & 37.3 & 4.2  & 13.1 \\
        (f). BASE & Noisy & Noisy & 4.4  & 33.3 & 4.1  & 10.3 \\
        \bottomrule
    \end{tabular}
    }
\end{table}

Concerning SNR, the WER is reduced the most in low SNR settings, i.e., high noise, as is shown in Figure~\ref{fig:overall-comparison}. This observation matches our hypothesis made in section~\ref{sec:pre-train-vs-scratch} about the domain discrepancy between pre-training and finetuning. Introducing noise during the pre-training stage bridges the domain gap and makes the model more resilient to noise at test time. One key takeaway from this work is that noise augmentation is needed during pre-training and finetuning phases to achieve the best AVSR performance in adverse acoustic conditions. Compared to training from scratch, the gain of noise-augmented pre-training peaks around 0dB SNR ratio, which matches the SNR used in training. 

Compared to ``\texttt{babble}'', ``\texttt{music}'' and ``\texttt{natural}'' noise types, noise-augmented AV-HuBERT pre-training is more effective in overlapping ``\texttt{speech}'' noise as shown in Figure~\ref{fig:overall-comparison}.  Consistent with previous findings when comparing audio-visual and audio-only recognizers trained from scratch, the visual modality provides a strong clue to ``choose'' the target speech track. AV-HuBERT effectively learns visual representation from paired audio-visual data, leading to significant gains in speech separation.

The gains from noise-augmented pre-training generalize across model architectures, as shown in table~\ref{tab:avg-base-comparison}. In addition, regardless of the finetuning strategy, our proposed pre-training approach is helpful, as is shown by row (b) vs. row (c) and row (e) vs. row (f) in N-WER in Table~\ref{tab:avg-base-comparison}.
\section{Conclusion}
\label{sec:conclusion}
This paper presented a new state-of-the-art audio-visual speech recognition (AVSR) model based on the AV-HuBERT approach for multimodal speech representation learning. To our knowledge, this is the first attempt towards building an AVSR model using a large volume of unlabeled audio-visual speech data. Our audio-visual speech recognizer achieves high recognition accuracy and is robust to different noise categories even with a few hours of labeled data. With less than 10\% of labeled data, our model outperforms prior SOTA by $\sim 50\%$. Our future work includes applying audio-visual speech recognition in real-world low-resource and multilingual settings.

\bibliographystyle{IEEEtran}
\bibliography{mybib}

% \newpage
\section{Appendix}
\label{sec:appendix}

\subsection{Full results}
\label{sec:app-full-results}

Table~\ref{tab:av-full-table} and \ref{tab:av-base-table} shows the WERs of LARGE and BASE AV-HuBERT models under various noise types and SNR levels.

\begin{table*}[hbt]
    \centering
    \caption{\label{tab:av-full-table}Test WER (\%) of LARGE AV-HuBERT under different levels and types of noise. Lower is better. B: babble, S: speech, M: music, N: natural noise.}
    % \resizebox{\linewidth}{!}{
    \begin{tabular}{crrrrrrrrrrrr}
    \toprule
    % \multicolumn{9}{c}{30h labeled}\\
    % \cmidrule(r){2-5}\cmidrule(r){6-9}
    \multirow{1}{*}{SNR (dB)} & \multicolumn{4}{c}{PT: None} & \multicolumn{4}{c}{PT: Clean} & \multicolumn{4}{c}{PT: Noisy} \\
    \cmidrule(r){1-5}\cmidrule(r){6-9}\cmidrule(r){10-13}
     A,30h & B & S & M & N & B & S & M & N & B & S & M & N \\
    %  \cmidrule(r){1-5}\cmidrule(r){6-9}\cmidrule(r){10-13}\cmidrule(r){14-17}
    %  & \multicolumn{16}{c}{30h labeled} \\
      \cmidrule(r){1-5}\cmidrule(r){6-9}\cmidrule(r){10-13}
-10 & 103.1 & 95.7 & 83.4 & 77.8 & 100.7 & 101.9 & 66.1 & 58.2 & 99.6 & 77.3 & 50.5 & 45.2 \\
-5 & 90.3 & 86.9 & 65.9 & 61.0 & 87.5 & 91.4 & 39.4 & 36.0 & 69.3 & 51.2 & 21.5 & 21.5 \\
0 & 63.9 & 70.5 & 46.4 & 46.6 & 37.6 & 63.9 & 17.4 & 15.9 & 21.9 & 32.0 & 9.0 & 9.3 \\
5 & 43.2 & 52.2 & 35.2 & 34.3 & 11.8 & 25.4 & 8.2 & 7.6 & 9.0 & 19.7 & 5.9 & 5.8 \\
10 & 32.1 & 39.9 & 28.0 & 27.8 & 6.7 & 9.5 & 5.9 & 5.5 & 5.6 & 10.8 & 4.9 & 4.6 \\
clean & \multicolumn{4}{c}{20.6} & \multicolumn{4}{c}{4.3} & \multicolumn{4}{c}{3.8}\\
\midrule
     A,433h & B & S & M & N & B & S & M & N & B & S & M & N \\
    %  \cmidrule(r){1-5}\cmidrule(r){6-9}\cmidrule(r){10-13}\cmidrule(r){14-17}
    %  & \multicolumn{16}{c}{30h labeled} \\
      \cmidrule(r){1-5}\cmidrule(r){6-9}\cmidrule(r){10-13}
-10 & 100.7 & 95.3 & 64.7 & 55.6 & 98.2 & 94.3 & 47.4 & 39.3 & 97.5 & 81.7 & 40.8 & 36.5 \\
-5 & 82.3 & 79.7 & 38.2 & 34.0 & 65.6 & 73.8 & 18.7 & 17.2 & 62.3 & 56.2 & 15.3 & 14.9 \\
0 & 39.2 & 52.8 & 18.8 & 17.7 & 17.0 & 46.3 & 6.5 & 6.4 & 15.7 & 37.3 & 5.7 & 5.6 \\
5 & 17 & 28.4 & 10.7 & 10.6 & 5.3 & 22.9 & 3.0 & 3.4 & 5.1 & 19.0 & 3.1 & 3.1 \\
10 & 8.4 & 15.7 & 7.5 & 7.3 & 2.7 & 9.7 & 2.0 & 2.2 & 2.6 & 8.3 & 2.3 & 2.3 \\
clean & \multicolumn{4}{c}{4.7} & \multicolumn{4}{c}{1.5} & \multicolumn{4}{c}{1.6}\\
\midrule
     AV,30h & B & S & M & N & B & S & M & N & B & S & M & N \\
    %  \cmidrule(r){1-5}\cmidrule(r){6-9}\cmidrule(r){10-13}\cmidrule(r){14-17}
    %  & \multicolumn{16}{c}{30h labeled} \\
      \cmidrule(r){1-5}\cmidrule(r){6-9}\cmidrule(r){10-13}
-10 & 80.3 & 62.7 & 60.3 & 55.6 & 32.2 & 18.1 & 16.4 & 13.2 & 30.7 & 11.5 & 12.5 & 11.4 \\
-5 & 63.1 & 52.1 & 46.8 & 44.9 & 18.5 & 10.4 & 9.3 & 8.0 & 15.9 & 6.8 & 7.3 & 6.6 \\
0 & 45.1 & 42.7 & 36.2 & 35.0 & 8.7 & 6.6 & 5.6 & 5.2 & 7.3 & 5.0 & 4.9 & 4.7 \\
5 & 33.3 & 34.8 & 29.5 & 28.5 & 4.8 & 4.8 & 4.3 & 4.1 & 4.4 & 4.2 & 4.1 & 4.0 \\
10 & 27.2 & 29.4 & 25.4 & 25.2 & 3.7 & 4.0 & 3.7 & 3.7 & 3.9 & 3.9 & 3.6 & 3.7 \\
clean & \multicolumn{4}{c}{20.8} & \multicolumn{4}{c}{3.3} & \multicolumn{4}{c}{3.3}\\
\midrule
     AV,433h & B & S & M & N & B & S & M & N & B & S & M & N \\
    %  \cmidrule(r){1-5}\cmidrule(r){6-9}\cmidrule(r){10-13}\cmidrule(r){14-17}
    %  & \multicolumn{16}{c}{30h labeled} \\
      \cmidrule(r){1-5}\cmidrule(r){6-9}\cmidrule(r){10-13}
-10 & 60.2 & 26.5 & 29.5 & 24.0 & 30.0 & 15.9 & 13.8 & 10.3 & 28.4 & 11.4 & 9.9 & 9.4 \\
-5 & 33.1 & 15.1 & 14.9 & 12.4 & 15.2 & 7.5 & 6.4 & 5.4 & 13.4 & 4.6 & 4.8 & 4.6 \\
0 & 14 & 8.8 & 7.9 & 8.0 & 5.9 & 3.9 & 3.3 & 2.9 & 5.0 & 2.9 & 2.5 & 2.5 \\
5 & 6.8 & 6.2 & 5.0 & 5.3 & 2.7 & 2.4 & 2.1 & 2.2 & 2.6 & 2.2 & 1.9 & 1.9 \\
10 & 4.6 & 5.1 & 4.0 & 4.3 & 1.9 & 1.9 & 1.7 & 1.8 & 1.9 & 1.8 & 1.8 & 1.7 \\
clean & \multicolumn{4}{c}{3.5} & \multicolumn{4}{c}{1.4} & \multicolumn{4}{c}{1.4}\\
        \bottomrule
    \end{tabular}
    % }
\end{table*}

\begin{table*}[hbt]
    \centering
    \caption{\label{tab:av-base-table}Test WER (\%) of BASE AV-HuBERT fine-tuned with 30 hours of labeled data under different levels and types of noise. Lower is better. B: babble, S: speech, M: music, N: natural noise.}
    % \resizebox{\linewidth}{!}{
    \begin{tabular}{crrrrrrrrrrrr}
    \toprule
    % \multicolumn{9}{c}{30h labeled}\\
    % \cmidrule(r){2-5}\cmidrule(r){6-9}
    \multirow{1}{*}{SNR (dB)} & \multicolumn{4}{c}{PT: None} & \multicolumn{4}{c}{PT: Clean} & \multicolumn{4}{c}{PT: Noisy} \\
    \cmidrule(r){1-5}\cmidrule(r){6-9}\cmidrule(r){10-13}
     A,30h (Clean-FT) & B & S & M & N & B & S & M & N & B & S & M & N \\
    %  \cmidrule(r){1-5}\cmidrule(r){6-9}\cmidrule(r){10-13}\cmidrule(r){14-17}
    %  & \multicolumn{16}{c}{30h labeled} \\
      \cmidrule(r){1-5}\cmidrule(r){6-9}\cmidrule(r){10-13}
-10 & 111.7 & 101.1 & 93.8 & 91.3 & 99.4 & 99.5 & 74.5 & 65.7 & 96.4 & 88.1 & 53.6 & 47.7 \\
-5 & 110.5 & 98.0 & 88.3 & 84.1 & 92.9 & 91.5 & 51.9 & 45.1 & 74.8 & 66.0 & 27.5 & 26.5 \\
0 & 99.8 & 92.4 & 76.7 & 72.7 & 57.3 & 74.2 & 25.5 & 23.7 & 29.7 & 48.0 & 12.3 & 12.6 \\
5 & 79.3 & 81.3 & 58.7 & 55.6 & 21.2 & 38.7 & 11.2 & 12.4 & 12.1 & 31.4 & 7.7 & 7.5 \\
10 & 52.6 & 60.8 & 43.5 & 44.0 & 9.4 & 17.6 & 7.4 & 7.5 & 7.1 & 15.3 & 6.1 & 5.9 \\
clean & \multicolumn{4}{c}{24.6} & \multicolumn{4}{c}{4.6} & \multicolumn{4}{c}{4.4}\\
\midrule
     A,30h (Noisy-FT) & B & S & M & N & B & S & M & N & B & S & M & N \\
    %  \cmidrule(r){1-5}\cmidrule(r){6-9}\cmidrule(r){10-13}\cmidrule(r){14-17}
    %  & \multicolumn{16}{c}{30h labeled} \\
      \cmidrule(r){1-5}\cmidrule(r){6-9}\cmidrule(r){10-13}
-10 & 103.1 & 94.3 & 80.2 & 75.3 & 107.0 & 92.2 & 65.6 & 55.8 & 98.6 & 87.5 & 53.2 & 46.8 \\
-5 & 89.4 & 86.3 & 61.7 & 55.9 & 81.2 & 74.9 & 38.9 & 31.8 & 74.5 & 65.1 & 26.6 & 24.5 \\
0 & 59.2 & 68.0 & 40.7 & 39.5 & 35.8 & 43.6 & 17.3 & 16.3 & 28.5 & 47.1 & 11.8 & 11.9 \\
5 & 38.3 & 48.6 & 28.8 & 29.0 & 13.8 & 20.7 & 9.1 & 9.8 & 11.3 & 31.1 & 7.1 & 7.5 \\
10 & 26.6 & 35.3 & 22.9 & 24.8 & 7.5 & 10.3 & 6.8 & 6.7 & 6.7 & 14.8 & 5.8 & 5.6 \\
clean & \multicolumn{4}{c}{16.9} & \multicolumn{4}{c}{4.8} & \multicolumn{4}{c}{4.4}\\
\midrule
     AV,30h (Clean-FT) & B & S & M & N & B & S & M & N & B & S & M & N \\
    %  \cmidrule(r){1-5}\cmidrule(r){6-9}\cmidrule(r){10-13}\cmidrule(r){14-17}
    %  & \multicolumn{16}{c}{30h labeled} \\
      \cmidrule(r){1-5}\cmidrule(r){6-9}\cmidrule(r){10-13}
-10 & 103.5 & 97.0 & 89.6 & 85.8 & 84.8 & 92.4 & 44.7 & 33.2 & 48.5 & 29.3 & 21.3 & 17.5 \\
-5 & 99.7 & 92.4 & 81.3 & 76.5 & 49.6 & 75.0 & 24.0 & 18.0 & 24.7 & 12.9 & 11.1 & 10.1 \\
0 & 86.6 & 83.5 & 66.2 & 60.6 & 19.7 & 41.1 & 11.2 & 9.3 & 11.4 & 7.7 & 6.9 & 6.7 \\
5 & 65.8 & 69.0 & 49.2 & 45.9 & 8.7 & 15.6 & 6.5 & 6.5 & 6.3 & 5.8 & 5.1 & 5.1 \\
10 & 42.2 & 50.5 & 35.7 & 36.1 & 5.5 & 7.8 & 5.0 & 5.2 & 4.7 & 5.0 & 4.4 & 4.6 \\
clean & \multicolumn{4}{c}{22.0} & \multicolumn{4}{c}{4.0} & \multicolumn{4}{c}{4.1}\\
\midrule
     AV,30h (Noisy-FT) & B & S & M & N & B & S & M & N & B & S & M & N \\
    %  \cmidrule(r){1-5}\cmidrule(r){6-9}\cmidrule(r){10-13}\cmidrule(r){14-17}
    %  & \multicolumn{16}{c}{30h labeled} \\
      \cmidrule(r){1-5}\cmidrule(r){6-9}\cmidrule(r){10-13}
-10 & 79.4 & 61.1 & 57.5 & 54.7 & 42.1 & 27.2 & 23.6 & 18.3 & 38.5 & 15.8 & 17.9 & 15.1 \\
-5 & 60.6 & 49.4 & 42.9 & 40.0 & 25.4 & 17.0 & 13.3 & 11.1 & 21.2 & 9.6 & 10.0 & 8.9 \\
0 & 42.0 & 38.8 & 32.4 & 30.7 & 12.8 & 10.0 & 7.9 & 7.4 & 9.9 & 6.7 & 6.5 & 6.2 \\
5 & 28.9 & 30.3 & 25.6 & 25.1 & 7.0 & 6.8 & 5.8 & 5.6 & 5.9 & 5.5 & 5.0 & 5.2 \\
10 & 22.8 & 24.4 & 21.4 & 21.3 & 5.3 & 5.2 & 4.9 & 4.8 & 4.8 & 4.8 & 4.3 & 4.5 \\
clean & \multicolumn{4}{c}{17.2} & \multicolumn{4}{c}{4.2} & \multicolumn{4}{c}{4.1}\\
        \bottomrule
    \end{tabular}
    % }
\end{table*}

\end{document}